# Strange Mass Corrections to HyperonicSemi-leptonicDecay in Statistical Model

[1]A. Upadhyay,[1]M. Batra

[1]School of Physics and Material Science, Thapar University, Patiala

## Abstract

The spin distributions, weak decay matrix elements for strange baryon octets with SU(3) breaking effects is studied. We systematically apply operator formalism along with statistical method to study $J^P = \frac{1}{2}^+$ strange baryon octets for their low energy properties. Baryon is taken as an ensemble of quark-gluon Fock states in sea with three valence quarks to have spin-1/2, color-1 and flavor-8 quantum numbers. Detailed balance principle is applied to calculate the probabilities of each Fock states, with the inclusion of mass correction of strange quark in order to check the SU(3) breaking in weak decays constants and spin distributions. A dominant contribution from the vector sea is verified as compared to scalar and tensor sea, also the symmetry breaking correction leads to the deviations in the value of axial vector matrix elements ratio $F/D$ from experimental values by 17%. Present framework suggests a stronger base to choose statistical model with detailed balance principle to verify the experimental and theoretical values available and hence provide a deeper understanding to the strange baryon structure. Symmetry breaking effects lead to reduction in the values of axial matrix elements. Its contribution has a significant role in determining the validity of present approach.

**Keywords:** Statistical Model, Detailed Balance, Strange baryons, Semi-Leptonic decays, SU(3)symmetry and its breaking.

**PACS:**14.20.Jn, 14.40.-n, 13.30.Ce, 11.30.HvEquation Section (Next)

## 1.1 Equation Chapter (Next) Section 1Introduction

The discovery of strange particles brought a new additive quantum number to the hadron physics. These particles appear in pairs during proton-proton collision and had unexpected large life-time. In order to have a deeper understanding for baryon structure at low energy, it becomes substantially important to have a detailed knowledge of their quark gluon contents. Low energy properties provide useful information about the detailed structure of baryon as well as other hyperons spectrum. In spite of hard efforts to study baryon spectra, we still lack to have complete information about several resonances observed at experiments. Unstable nature of these particles and non-perturbative aspect of QCD put constrains on further understanding. Phenomenological models [1-2] revealed information regarding various hyperonic states. Recently, the lowest lying baryonic states have been studied by lattice QCD collaborations [3-4], but there is less consistency among different outcomes. During last 30 years, efforts have been made to solve the proton spin problem [5]. In the framework of QCD, quarks interact through color forces mediated by vector gluons. Existence of quark gluon interaction implies that $\bar{q}q$ pairs can be created by the virtual gluons emitted from valence quarks. Now, it is quite well known that the spin distribution comes from gluons and orbital angular momentum of quarks and confirms the sea quark-gluon structure of hadrons. Chiral symmetry for strong interaction is almost exact in the light of u and d flavor sector, but it becomes approximate when strangeness is also included in sea, due to the large mass of strange quark. In standard model SU(3) symmetry breaking is given by current quark mass term in QCD Lagrangian with $m_{u,d} << m_s$. At zero order $m_s$, semileptonic decays and spin distribution are written in terms of F and D, that are completely SU(3) symmetric. However, at first order $m_s$ one expects a small SU(3) symmetry breaking in the theory due to non-zero mass of strange quark. It also comes from mismatch of quark's wave function and recoil effects. Yang et al. [6] studied the hyperon polarization effects and claimed that the experimental data on hyperon polarization seems to favor the theoretical predictions of SU(3) symmetry breaking. Chiral Quark Soliton Model studied breaking in SU(3) by describing baryons as rotating solitons adiabatically in flavor space, the model uses a collective Hamiltonian term which includes the flavor rotation degrees of freedom and strange mass correction linear in $M_s$[7]. On the other hand, chiral constituent quark model in ref. [8] adds the $m_s$ contribution to weak decay ratios by considering the Goldstone boson masses as non-degenerate. Various phenomenologists have suggested different models[9-15] to extract $V_{us}$ from semi-leptonic decays but few models give data in disagreement with other available results. High precision measurement of axial vector coupling ratios like $\Sigma^- \to n$ [16] and $\Lambda \to p$ [17] demands an approach for analysis of SU(3) breaking in detail. In fact the ratio $g_1/f_1$ in the limit of SU(3) breaking matches with the latest data of NA48/1 Collaboration [18].

The statistical approach finds the individual contributions from strange and non-strange components of sigma and cascade baryonic systems. Assuming that the hadrons can be expanded in terms of quark-gluon Fock states, we need to incorporate the effects of strange quark mass into the probabilities associated with these Fock states. Thus we chose a frame-work where the constrains due to non-negligible mass of strange quark in terms of the state densities is calculated from of detailed balance principle [22].Statistically, we can think of arbitrary number of gluons to be present in the sea, but due to limited free energy of gluon and suppressed number of strange quark-antiquark pair, the total numbers of partons are restricted to six. In present article, Section 1.2 describes the principle of detailed balance and the statistical approach for $J^P = \frac{1}{2}^+$ strange baryons having various quark-gluon Fock states including strange quark condensates. Section 1.3 gives a brief discussion of SU(3) symmetry breaking analysis in various static low energy properties of strange baryons. A unique combination of possible states forming baryon wave function with strange mass corrections in valence and sea part leads to a justified analysis of the quark dynamics. Section 1.4 discusses the results and graph followed by a conclusion in section 1.5.Equation Section (Next)

## 1.2 Detailed Balance principle with Mass Corrections:

The article is based on calculation of certain multiplicities in spin and color space for different possible sets of the Fock states like $|\overline{uu}g\rangle, |\overline{dd}g\rangle, |gg\rangle, |\overline{uu}\overline{ss}\rangle$ etc. We suggest a possible way to construct the hadronic wave function, so that every Fock state has a definite probability in flavor, spin and color space and maintain the antisymmetrization of the total function. We apply detailed balance principle to the strange baryons octet where strange quark antiquark pairs splits and recombine and they have dependence over the number of gluons already present in sea part which cannot be ignored. The principle [23-24] considers equilibrium between splitting and recombination of any of the two Fock states so that the ratio of rates of transitions between any of the two Fock states is equal to ratio of their densities. Thus

$$\frac{R_A}{R_B} = \frac{\rho_A}{\rho_B}$$
(1.2.1)

A suitable density operator comprising of densities of all Fock states can be defined as:

$$\bar{\rho} = \sum_{i,j,l,k} \rho_{i,j,l,k} |\{qqq\},\{i,j,l,k\}\rangle\langle\{qqq\},\{i,j,l,k\}|$$
(1.2.2)

Where $\rho_{i,j,k,l}$ is the probability of finding a specific Fock state with strange quarks where i,j and l are the numbers of, $\overline{uu}$, $\overline{dd}$ and $\overline{ss}$ respectively and k are the number of gluons. To calculate the associated probability of each Fock state, three different sub-processes ($g \Leftrightarrow q\overline{q}$, $g \Leftrightarrow gg$ and $q \Leftrightarrow qg$) are taken into consideration. Although the transition processes may depend upon many factors in reality like mass, parton numbers and quantum numbers like spin, flavor and color of the partons but for simplicity we firstly take into account the parton-number of the partons involved. Another constraint is due to unavoidable mass of strange quark which affects the transition probability. Flavor probability is carried forward to find the different multiplicities in spin and color space with suitable combinations.

(i) When $q \Leftrightarrow qg$ is considered: The general expression [39] of probability for the same subprocess is written as:

$$|\{q\},\{i,j,l,k-1\}\rangle \underset{(3+2i+2j+2l)k}{\overset{3+2i+2j+2l}{\Leftrightarrow}} |\{q\},\{i,j,l,k\}\rangle$$
(1.2.3)

Where i refer to $\overline{uu}$ pairs, j refer to $\overline{dd}$, l refers to $\overline{ss}$ and k refers to no. of gluons so that total number of partons are 3+2i+2j+2k+2l in the final state.

$$\frac{\rho_{i,j,l,k}}{\rho_{i,j,l,k-1}} = \frac{1}{k}$$

(1.2.4)

(ii) When both the processes $g \Leftrightarrow gg$ and $q \Leftrightarrow qg$ are included: Similarly,

$$\left| \{q\}, \{i,j,l,k-1\} \right\rangle \xrightarrow[\frac{3+2i+2j+2l+k-1}{(3+2i+2j+2l)k+\frac{k(k-1)}{2}}]{} \left| \{q\}, i,j,l,k \right\rangle$$

(1.2.5)

$$\frac{\rho_{i,j,l,k}}{\rho_{i,j,l,k-1}} = \frac{(3+2i+2j+2l+k-1)}{(3+2i+2j+2l)k + \frac{k(k-1)}{2}}$$

(1.2.6)

(iii) When $g \Leftrightarrow q\bar{q}$ is considered:- In this case, the sub-processes involving $g \rightleftharpoons s\bar{s}$, makes the situation critical. The non-zero mass of strange quark constrains the free energy of gluon and suppress the number of $s\bar{s}$ pairs in sea in agreement with ref. [38].The mass correction imposed to a flavor asymmetric sea leads to a smaller probability for Fock states having strange quark. If n is the total number of partons then the constraint appear in the form of $k(1-C_i)^{n-1}$ [23] and in all cases, $C_{l-1} = \frac{2M_s}{M_B - 2(l-1)M_s}$, Ms is the mass of s-quark and $M_B$ is the mass of strange baryon.

$$\left| \{q\}, g \right\rangle \xrightarrow[\frac{1X(1-C_0)^3}{2X1}]{} \left| \{q\}, s\bar{s} \right\rangle$$

$$\left| \{q\}, u\bar{u}g \right\rangle \xrightarrow[\frac{1(1-C_0)^3}{1X2}]{} \left| \{q\}, u\bar{u}s\bar{s} \right\rangle$$

Generalizing it to' k' no. of gluons and 'l' number of strange quark-antiquark pairs for a singly strange baryon.

$$\left\{ \left| \{q\}, i,j,l-1,k \right\rangle \right\} \xrightarrow[\frac{k(1-C_{l-1})^{n-1}}{l(l+1)}]{} \left\{ \left| \{q\}, i,j,l,k \right\rangle \right\}$$

(1.2.7)

$$\frac{\rho_{i,j,l,k}}{\rho_{i,j,l-1,k}} = \frac{k(1-C_{l-1})^{n-1}}{l(l+1)}$$

(1.2.8)

Suppose no $s\bar{s}$ pair is present initially and generation of one pair requires gluon to have sufficient energy and the condition becomes:

$$\left| \{q\}, i,j,0,k \right\rangle \xrightarrow[\frac{k(1-C_0)^{n-2l-1}}{1(l+1)}]{} \left| \{q\}, i,j,1,k-1 \right\rangle \text{ where } n = 3+2i+2j+2l+k$$

$$\frac{\rho_{i,j,1,k-1}}{\rho_{i,j,0,k}} = \frac{k(1-C_0)^{n-2l-1}}{1(l+1)}$$

(1.2.9)

$$\left|\{q\},i,j,1,k-1\right\rangle \xrightleftharpoons[\frac{(k-1)(1-C_1)^{n-2l}}{2(2+1)}]{} \left|\{q\},i,j,2,k-2\right\rangle$$

$$\frac{\rho_{i,j,2,k-2}}{\rho_{i,j,1,k-1}} = \frac{(k-1)(1-C_1)^{n-2l}}{2(2+1)}$$

(1.2.10)

Here the go-out probability depends upon the number of partons present at that time. Similarly we proceed until all gluons have been converted into strange pairs.

$$\left|\{q\},i,j,k-1,1\right\rangle \xrightleftharpoons[\frac{1(1-C_{k-1})^{n-k+2}}{k(k+1)}]{} \left|\{q\},i,j,k,o\right\rangle$$

(1.2.11)

$$\left|\rho_{i,j,k,0}\right\rangle \xrightleftharpoons[\frac{1(1-C_{k-1})^{n-k+2}}{k(k+1)}]{} \left|\rho_{i,j,k-1,1}\right\rangle$$

Thus

$$\frac{\rho_{i,j,0}}{\rho_{i,j,0,l}} = \frac{(k(k-1)(k-2)(k-3)---1(1-C_0)^{n-2l-1}(1-C_1)^{n-2l}(1-C_2)^{n-2l+1}(1-C_{l-1})^{n-k-2})}{k!(k+1)!}$$

(1.2.12)

$$\left|\{q\},i,j,l+k-1,1\right\rangle \xrightleftharpoons[\frac{1(1-C_{l+1})^{n-k+2}}{l+k(l+k+1)}]{} \left|\{q\},i,j,l+k,0\right\rangle.$$

Generalizing it to any number of gluons, the ratio becomes:

$$\frac{\rho_{i,j,k}}{\rho_{i,j,l+k,0}} = \frac{(k(k-1)(k-2)(k-3)---1(1-C_0)^{n-2l-1}(1-C_1)^{n-2l}(1-C_2)^{n-2l+1}---(1-C_{l-1})^{n+k-2})}{(l+1)(l+2).......(l+k)(l+k+1)}$$

(1.2.13)

Complete expressions for sigma triplet and cascade doublet can be written as:

| Baryon | Quark Content | Expression |
|--------|---------------|------------|
| $\Sigma^+$ | Uus | $\dfrac{\rho_{i,j,l+k,0}}{\rho_{0,0,0,0}} = \dfrac{2}{i!i+1!j!(j+1)!(l+k)!(l+k+1)!}$ |
| $\Sigma^-$ | Dds | $\dfrac{\rho_{i,j,l+k,0}}{\rho_{0,0,0,0}} = \dfrac{2}{i!i!j!(j+2)!(l+k)!(l+k+1)!}$ |
| $\Sigma^0$ | Uds | $\dfrac{\rho_{i,j,l+k,0}}{\rho_{0,0,0,0}} = \dfrac{1}{i!(i+1)!j!(j+1)!(l+k)!(l+k+1)!}$ |
| $\Xi^0$ | Uss | $\dfrac{\rho_{i,j,l+k,0}}{\rho_{0,0,0,0}} = \dfrac{2}{i!i+1!j!(j)!(l+k)!(l+k+2)!}$ |
| $\Xi^-$ | Dss | $\dfrac{\rho_{i,j,l+k,0}}{\rho_{0,0,0,0}} = \dfrac{2}{i!i!j!(j+1)!(l+k)!(l+k+2)!}$ |

Thus, all the probabilities can be written in terms of $\rho_{0,0,0,0}$ which can be solved by applying the constraint that total probability is one. This all sets of probabilities for each Fock states can be obtained.

## 1.3 Equation Section (Next)Statistical Framework:

The statistical approach assumes the statistical decomposition of the baryonic state in various quark-gluon Fock states in which the relevant operators act on the sea part as well [21,32]. To show an active participation of sea, a suitable wave-function with valence and non-valence quark gluon Fock states are written so as to get total a spin-1/2, color singlet, and flavor octet baryons[20]. For the lowest lying baryons in S-wave, we assume a flavorless sea with suitable spin (0,1,2) and color ($1_c$, $8_c$, $10_c$) combination. The valence and sea contents are suitably arranged to maintain the anti-symmetry of the total wave-function $\Psi = \Phi(|\phi\rangle|\chi\rangle|\psi\rangle|\xi\rangle)$ .Let $\varphi_1^{1/2}$ is the standard SU(3) $q^3$ wave-function transforming as 56 of SU(6) wave-function:

$$\left|\Phi_{\frac{1}{2}}^{\uparrow}\right\rangle = \frac{1}{N}[\phi_1^{(\frac{1}{2})^{\uparrow}} H_0 G_1 + a_8 \phi_8^{(\frac{1}{2})^{\uparrow}} H_0 G_8 + a_{10} \phi_{10}^{(\frac{1}{2})^{\uparrow}} H_0 G_{\overline{10}} + b_1 [\phi_1^{\frac{1}{2}} \otimes H_1]^{\uparrow} G_1 + b_s (\phi_8^{\frac{1}{2}} \otimes H_1)^{\uparrow} G_8 +$$
$$b_{10} (\phi_{10}^{\frac{1}{2}} \otimes H_1)^{\uparrow} G_{\overline{10}} + c_8 (\phi_8^{\frac{3}{2}} \otimes H_1)^{\uparrow} G_8 + d_8 (\phi_8^{\frac{3}{2}} \otimes H_2)^{\uparrow} G_8]$$

(1.3.1)

Where
$$N^2 = 1 + a_8^2 + a_{10}^2 + b_1^2 + b_8^2 + b_{10}^2 + c_8^2 + d_8^2$$

The above wave-function includes seven coefficients which are to be determined statistically. $H_{0,1,2}$ & $G_{1,8,10}$ denotes the spin and color possibilities of sea quark-gluon Fock states. The first three terms in the wave-function describe a spin ½ $q^3$ state coupled to a spin(0) scalar sea and the other terms with co-efficients ($b_1, b_8, b_{\overline{10}}, c_8$) describe a spin ½ $q^3$ state coupled to spin 1(vector sea) where $d_8$ is signifying a tensor contribution. The above mentioned wave-function can also be rewritten in the form of $\phi_{val}\phi_{sea}$ and the coefficients ($a_0, a_8, a_{\overline{10}}, b_1, b_8, b_{\overline{10}}, c_8, d_8$) by a factor $\sum n_{\mu\nu} * c_{sea}$ in the wave-function $\left|\Phi_{\frac{1}{2}}^{\uparrow}\right\rangle = \sum_{\mu,\nu} (n_{\mu\nu} * c_{sea}) \phi_{val}\phi_{sea}$ Here μ and ν have values 0, 1, 2 and 1, 8, $\overline{10}$ respectively. It leads to nine terms in the wave-function but μ=2 combines only with ν=8 in a view to maintain the anti-symmetrization of the wave-function. This means that, in spite of nine terms we are left with seven terms in the wave-function with seven coefficients. All $n_{\mu,\nu}'s$ are calculated from multiplicities of each Fock state in spin and color space. These multiplicities are expressed in the form of $\rho_{j_1,j_2}$ where the core quark part has spin $j_1$ and the sea quark-gluons carry spin $j_2$ so that resultant spin is ½. Such a comparison of multiplicities in Fock space serves two purposes. The first one is to find a common multiplier ("c") for each particular combination of valence and sea. The second is to calculate probability of each substate with specified specific spin, color quantum numbers. For instance, a simple two gluon sea can have spin 0,1,2 and color as 1,8, $\overline{10}$ and similarly for the higher number of gluons. A single gluon Fock state in the sea being spin 1 and color octet state will contribute to only $H_1G_8$ component of sea. Each coefficient will have a particular value of $\sum n_{fv} c_{sea}$ combination depending upon the Fock content

$$a_8 = (n_{08}c_{sea})_{|gg\rangle} + (n_{08}c_{sea})_{|\overline{u}ug\rangle} + (n_{08}c_{sea})_{|\overline{d}dg\rangle} + (n_{08}c_{sea})_{|\overline{s}sg\rangle} + \ldots.$$

$$b_1 = (n_{11}c_{sea})_{|gg\rangle} + (n_{11}c_{sea})_{|\overline{u}ug\rangle} + (n_{11}c_{sea})_{|\overline{d}dg\rangle} + (n_{11}c_{sea})_{|\overline{s}sg\rangle} + \ldots\ldots$$

$$b_{\overline{10}} = (n_{1\overline{10}}c_{sea})_{|gg\rangle} + (n_{1\overline{10}}c_{sea})_{|\overline{u}ug\rangle} + (n_{1\overline{10}}c_{sea})_{|\overline{d}dg\rangle} + (n_{1\overline{10}}c_{sea})_{|\overline{s}sg\rangle} + \ldots\ldots$$

In a similar way, we can get all the coefficients in the expansion of baryonic state. Each baryon octet contains set of possible Fock states ($|gg\rangle, |\overline{u}ug\rangle, |\overline{d}dg\rangle, |\overline{s}sg\rangle, |\overline{u}u\overline{d}d\rangle, |\overline{u}u\overline{s}s\rangle, |\overline{u}ugg\rangle, |\overline{d}dgg\rangle$ etc.) but their total probability differs due to the mass inherited from the flavor probability ratio. The other factor affecting the overall probabilities comes through the normalization procedure. All variation leads to different coefficients for each baryon wave function used in the expansion of baryon octets. To calculate the physical properties, it is convenient to write contributions from scalar, vector and tensor sea in the form of two parameters $\alpha$, $\beta$ [20]:

$$\alpha = \frac{1}{N^2}\frac{4}{9}(2a + 2b + 3d + \sqrt{2}e) = \frac{2(6 + 3a_8^2 - 2b_1^2 - b_8^2 + 4b_8c_8 + 5c_8^2 - 3d_8^2)}{27(1 + a_{10}^2 + a_8^2 + b_1^2 + b_{10}^2 + b_8^2 + c_8^2 + d_8^2)} \qquad (1.3.2)$$

$$\beta = \frac{1}{9N^2}(2a - 4b - 6c - 6d + 4\sqrt{2}e) = \frac{3 - 9a_{10}^2 - 3a_8^2 - b_1^2 + 3b_{10}^2 + b_8^2 + 8b_8c_8 - 5c_8^2 + 3d_8^2}{27(1 + a_{10}^2 + a_8^2 + b_1^2 + b_{10}^2 + b_8^2 + c_8^2 + d_8^2)}$$

$$(1.3.3)$$

Where a, b, c, d and e are defined as: $a = \frac{1}{2}(1 - \frac{b_1^2}{3}), b = \frac{1}{4}(a_8^2 - \frac{b_8^2}{3}), c = \frac{1}{2}(a_{10}^2 - \frac{b_{10}^2}{3}), d = \frac{1}{18}(5c_8^2 - 3d_8^2), e = \frac{\sqrt{2}}{3}(b_8c_8)$

All the relevant properties are written in terms weak decay matrix elements F and D which are calculated from the above values of $\alpha$ and $\beta$ that signifies the individual coefficients of each Fock states. The importance of $\alpha$ and $\beta$ lies in the fact that these parameters can directly be related to the spin distribution of quarks inside the baryonic system. To check the consistency of our previous approach [32] with the present technique we study the contributions from tensor, scalar and vector part of the total sea.

## 1.4 SU(3) Analysis for Static properties of Strange Baryon:

Various studies [7-14, 19,34,36] on SU(3) breaking for the matrix elements F and D, suggest importance of strange mass corrections in the static properties of baryon. Avenarious [14] suggested that SU(3) symmetry breaking in constituent quark level give F/D=0.73±0.09 , which is larger than that of SU(3) symmetric value(0.59±0.02). Ratcliffe [18] using weak decay ratios compared the values of F and D in both the cases and claimed that the difference of these two values is found to be 0.012. In order to have an accurate prediction of matrix elements F and D, an elaborate data related to the individual contribution of valence and sea quark gluon Fock state is needed. In light of above investigations, we studied the SU(3) symmetry breaking effects using statistical approach. The discussion given below is associated with both $\Delta s = 0$ and $\Delta s = 1$ decays. The first moment of spin structure function $g_1^p$ [7] of proton is given by

$$I_1^p = \int_0^1 dx g_1^p(x) = \frac{1}{18}(4\Delta u_p + \Delta d_p + \Delta s_p)(1 - \frac{\alpha_s}{\pi}) \qquad (1.4.1)$$

We should note that $\alpha_s$ is defined from Bjorken sum rule and different authors [25-26] use different fits for the same purpose as ours. The current operator for axial-vector is: $J_\mu^\sigma = \bar{\psi}\gamma_\mu\gamma_5\frac{\lambda^\sigma}{2}\psi$, where σ =0 to 8 with $\psi$ as the quark field triplet and $\bar{\psi}$ as the conjugate quark field. The baryon states are constructed with each matrix element of the octet axial currents for $B_j$ and $B_k$ so that $\langle B_j | J_\mu^\sigma | B_k \rangle = Dtr(J_\mu^i\{B_k, \overline{B_j}\}) + Ftr(J_\mu^i[B_k, \overline{B_j}])$. F and D couplings are proportional to structure constant $f_{ijk}$ and symmetric invariant tensor $d_{ijk}$. Neglecting higher order terms we obtain $\Gamma_1^p = (4\Delta u_p + \Delta d_p + \Delta s_p)$ and $g_A/g_V = F + D$ for proton, where the reduced matrix elements obtained from the decay amplitudes of hyperonic semi-leptonic decays. Philip G. Ratcliffe [35] examined the data of hyperonic decays and parameterized the matrix parameters F and D. They calculated the deviation due to symmetry breaking to be just 2%. Different authors suggest change in the new fitted values of F and D but Cheng and Li [36] mentioned that there exists no priori reason to expect the corrections to either increased or decreased ratio.

The experimental values for F and D come out to be 0.46 and 0.80 respectively [24]. From the available $\Delta\Sigma$ from DIS experiments and estimated F and D, all other values can be predicted for all baryon octet. Similarly, $g_A/g_V(\Sigma^- \rightarrow n) = F - D$ and other decay ratios can also be defined in terms of SU(3) symmetric F and D. For calculating the properties, an operator $\hat{o} = \sum_i \hat{O}_f^i \hat{\sigma}_z^i$ is to be defined where $\hat{o}_f^i$ depends upon on the flavor of i$^{th}$ quark and $\hat{\sigma}_z^i$ is the spin projection operator of i$^{th}$ quark. We employ symmetry breaking effects both in sea and valence. Flavor symmetry in sea is said to be broken by the mass difference between strange and non-strange light quarks. This leads to modifications in the probabilities of Fock states with strange quark-antiquark pairs. The constraints appear further due to the processes $g \rightleftharpoons s\bar{s}$ arising due to strange mass corrections. The gluon is restricted to possess sufficient free energy in order to produce a strange quark-antiquark pair with non-negligible mass. Thus the transition probabilities for $g \rightleftharpoons s\bar{s}$ involves two factor k and $C_1$ in the form of $k(1 - C_j)^{n-1}$ as discussed. Suitable mass corrections when implemented in strange sea, leads to suppression of contributions from various Fock states containing strange quark–antiquark pair. The same can be realized from the lower value of the co-efficients in table 3.1. The coefficients represent the individual contributions from various parts of sea categorized as scalar, vector and tensor depending upon their spin from the table, it is clear that contributions from various parts of sea get reduced except in case of spin one and color octet. Thus SU(3) breaking in sea favors spin one and color octet. This may be due to the lesser splitting and recombination between gluon and a strange quark-antiquark pair.

Conventionally, axial vector coupling ratios ($g_A/g_V$) are directly expressed in terms of weak decay matrix elements F and D. We express the symmetry breaking effects in valence part in terms of conventional parameters F and D and symmetry breaking parameter "r". The modified expressions in terms of axial matrix elements F, D and r is mentioned below:

$$g_A/g_{V(\Xi^- \to \Sigma^0)} = (\frac{2}{3} + \frac{r}{3})F + (\frac{2}{3} + \frac{r}{3})D$$

(1.4.2)

$$g_A/g_{V(\Xi^- \to \Lambda)} = \frac{1}{9}((6+5r)F + (2+r)D)$$

(1.4.3)

$$g_A/g_{V(\Sigma^- \to n)} = \frac{1}{3}(2+r)F - rD)$$

(1.4.4)

$$g_A/g_{V(\Xi^0 \to \Sigma^+)} = (\frac{2}{3} + \frac{r}{3})F + (\frac{2}{3} + \frac{r}{3})D$$

(1.4.5)

$g_A/g_V$ plotted against 'r' for possible range of r depending upon $m/m_s$ ratio where m is the average mass of u and d quarks. The experimental data of weak decay ratio suggests a feasible value of r. Best fit value of r is obtained by using a suitable fitting algorithm and found to be r=0.859. Best fit value of r will give the theoretical results for spin distribution, and weak decay constants provided we know the axial vector matrix elements calculated from the statistical approach using detailed balance principle. The results are discussed on the basis of SU(3) broken sea. This study makes a platform for analysis of low energy properties of strange baryons.

| Coefficients | $a_8$ | $a_{\overline{10}}$ | $b_1$ | $b_8$ | $b_{\overline{10}}$ | $c_8$ | $d_8$ |
|---|---|---|---|---|---|---|---|
| $\Sigma^+$ | 0.20 | 0.09 | 0.54 | 0.51 | 0.06 | 0.32 | 0.06 |
| | 0.18 | 0.06 | 0.36 | 0.58 | 0.04 | 0.41 | 0.04 |
| $\Sigma^0$ | 0.22 | 0.09 | 0.56 | 0.66 | 0.07 | 0.46 | 0.07 |
| | 0.17 | 0.06 | 0.35 | 0.74 | 0.04 | 0.52 | 0.04 |
| $\Sigma^-$ | 0.20 | 0.09 | 0.54 | 0.29 | 0.07 | 0.18 | 0.06 |
| | 0.19 | 0.07 | 0.42 | 0.30 | 0.05 | 0.21 | 0.05 |
| $\Xi^0$ | 0.20 | 0.09 | 0.53 | 0.16 | 0.06 | 0.16 | 0.06 |
| | 0.19 | 0.06 | 0.40 | 0.26 | 0.05 | 0.18 | 0.04 |
| $\Xi^-$ | 0.23 | 0.1 | 0.6 | 0.52 | 0.08 | 0.52 | 0.07 |
| | 0.19 | 0.07 | 0.42 | 0.5 | 0.05 | 0.50 | 0.046 |

Table 3.1 Seven parameters for $J^P = \frac{1}{2}+$ strange baryons octets in complete SU(3) symmetric sea in upper row and its breaking in lower row.

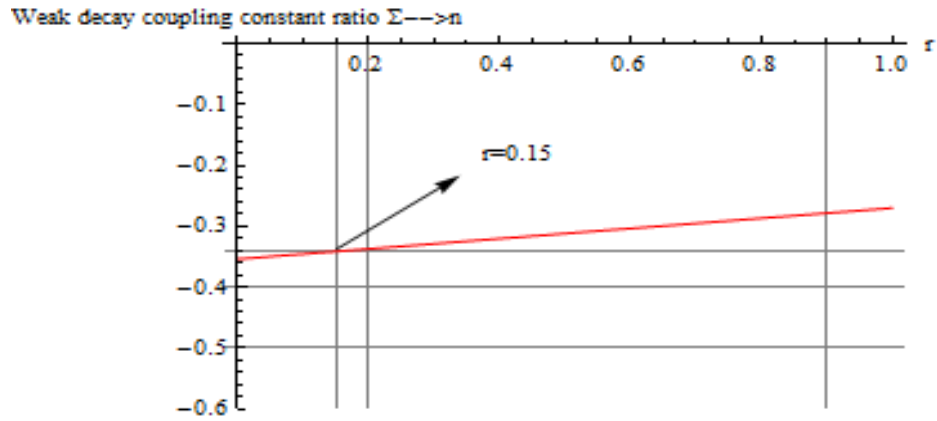

Weak decay coupling constant ratio $\Xi(-)\text{---}>\Lambda$

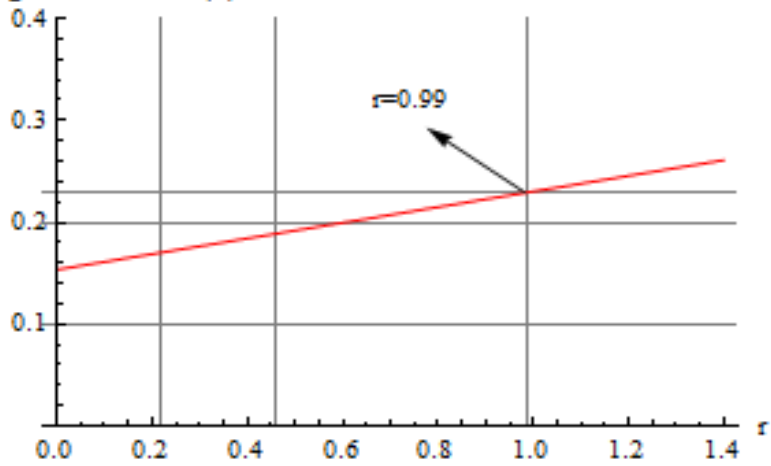

Weak decay coupling constant ratio $\Xi 0\text{-->}\Sigma+$

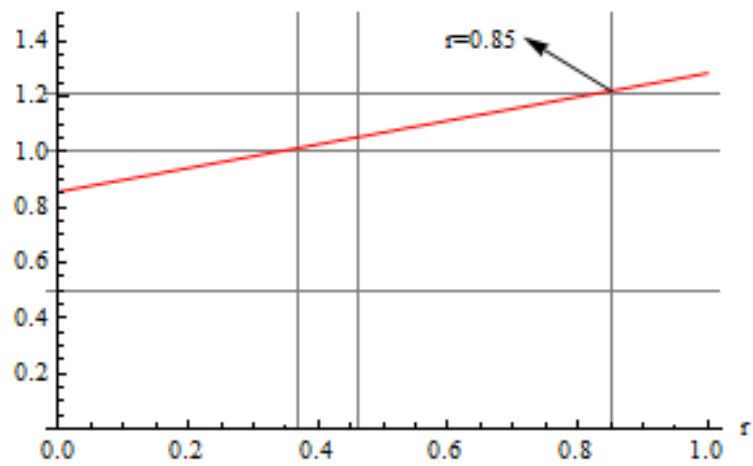

Weak decay coupling constant ratio $\Xi(-)\text{--->}\Sigma 0$

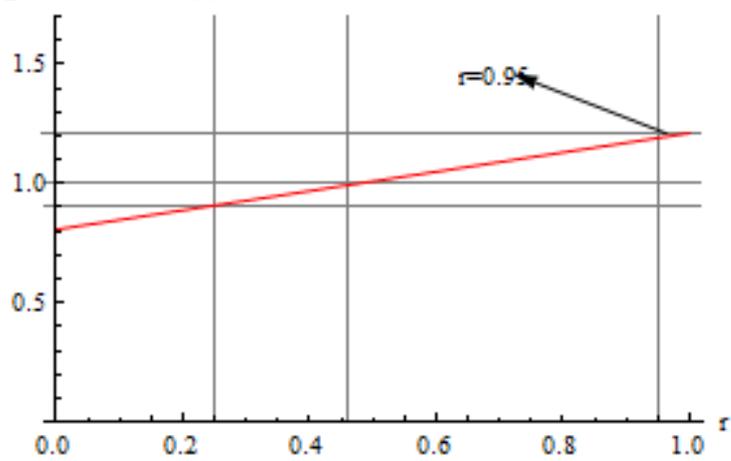

| Sr. No | Hyperonic weak decay ratios | Statistical Model (SU(3) symmetry) | Statistical Model (SU(3) Breaking in Sea) | Statistical Model(SU(3) breaking in (sea+valence) | Theoretical values [20] (SQM) | Chiral Quark Soliton Model | Expt. Result or PDG[29] | Chiral Constituent Quark Model [8] | |
|---|---|---|---|---|---|---|---|---|---|
| | | | | | | | | SU(3) symmetry | SU(3) breaking |
| 1. | $g_A/g_{V(\Xi^-\to\Sigma^0)}$ | 1.11 | 1.29 | 1.15 | ---- | 1.29[28] | ------ | 0.95 | 1.27 |
| 2. | $g_A/g_{V(\Xi^-\to\Lambda)}$ | 0.25 | 0.26 | 0.22 | 0.333 | 0.2115[27] 0.22[28] | 0.25±0.5 | 0.21 | 0.21 |
| 3. | $g_A/g_{V(\Sigma^-\to n)}$ | -0.22 | -0.28 | -0.27 | ----- | -0.3101[27] -0.31[28] | -0.34±0.17 | -0.16 | -0.31 |
| 4 | $g_A/g_{V(\Xi^-\to\Xi^0)}$ | -0.22 | -0.28 | -0.28 | -0.333 | -0.3101[27] | ------- | -0.16 | -0.31 |
| 5. | $g_A/g_{V(\Xi^0\to\Sigma^+)}$ | 1.11 | 1.29 | 1.22 | ---- | ----- | 1.21±0.05 | 0.95 | 1.27 |

Table 3.2- weak decay constant computed in statistical model with SU(3) symmetry and its breaking in sea.

| Strange Baryons (with quark content) | Integrated polarized quark densities | Statistical Model | | | Chiral Quark Soliton Model | |
|---|---|---|---|---|---|---|
| | | Exact SU(3) Symmetry | SU(3) breaking (sea) | SU(3) breaking(valence & sea) | Exact SU(3) Symmetry[27] | Symmetry Breaking[27] |
| $\Sigma^+$ (uus) | $\Delta u = \frac{1}{3}F(r+6) - \frac{1}{9}Dr$ | 0.81 | 0.87 | 0.99 | +0.98±0.023 | +0.73±0.17 |
| | $\Delta d = -\frac{1}{9}(D-3F)$ | 0 | 0 | -0.07 | -0.02±0.09 | -0.37±0.019 |
| | $\Delta s = \frac{1}{3}F(r+3) - \frac{1}{9}D(r+9)$ | -0.23 | -0.26 | -0.185 | -0.29±0.13 | -0.18±0.39 |
| $\Sigma^-$ (dds) | $\Delta u = -\frac{1}{9}r(D-3F)$ | 0 | 0.042 | 0.072 | -0.02±0.09 | -0.37±0.019 |
| | $\Delta d = \frac{1}{3}F(r+6) - \frac{1}{9}Dr$ | 0.88 | 0.91 | 1.07 | +0.98±0.23 | +0.73±0.17 |
| | $\Delta s = \frac{1}{3}F(r+3) - \frac{1}{9}D(r+9)$ | -0.22 | -0.28 | -0.18 | -0.29±0.13 | -0.18±0.39 |
| $\Sigma^0$ (uds) | $\Delta u = -\dfrac{Dr}{9} + \dfrac{Fr}{3} + F$ | 0.41 | 0.46 | 0.44 | +0.48±0.16 | +0.18±0.08 |
| | $\Delta d = -\dfrac{Dr}{9} + \dfrac{Fr}{3} + F$ | 0.41 | 0.46 | 0.44 | +0.48±0.16 | +0.18±0.08 |
| | $\Delta s = \dfrac{1}{3}F(r+3) - \dfrac{D}{9}(r+9)$ | -0.23 | -0.28 | -0.27 | -0.29±0.13 | -0.18±0.39 |
| $\Xi^0$ (uss) | $\Delta u = \dfrac{1}{3}F(r+3) - \dfrac{1}{9}D(r+9)$ | -0.21 | -0.25 | -0.15 | -0.29±0.13 | -0.14±0.21 |
| | $\Delta d = -\dfrac{1}{9}r(D-3F)$ | 0 | 0 | 0.008 | -0.02±0.09 | -0.37±0.19 |
| | $\Delta s = \dfrac{1}{9}(3(D+F(r+6)) - D(r+3))$ | 0.89 | 1.03 | 1.06 | 0.98±0.23 | 1.50±0.60 |
| $\Xi^-$ (dss) | $\Delta u = \dfrac{-1}{9}r(D-3F)$ | 0 | -0.06 | -0.07 | -0.02±0.09 | -0.37±0.19 |
| | $\Delta d = \dfrac{1}{3}F(r+3) - \dfrac{1}{9}D(r+9)$ | -0.19 | -0.18 | -0.17 | -0.29±0.13 | -0.14±0.21 |
| | $\Delta s = \dfrac{1}{9}(3F(r+6) - D(r-3))$ | 0.77 | 1.19 | 1.2 | 0.98±0.23 | 1.50±0.60 |

Table 3.3- Showing spin distribution of strange baryons in statistical model and comparison with other theoretical models.

## 1.5 Results:

In the present work, we focus on the semileptonic decays and axial vector coupling constant in SU(3) symmetry sea and its breaking using statistical approach. Here the sea is an active participant with direct inclusion of strange mass corrections and QCD one body operator directly involves the strange quark mass in the form of parameter "r". We provide a best fit for the axial vector matrix elements F and D and find the contribution of strange quark to the spin of proton.

Table 3.1 shows the contribution of vector, scalar and tensor parts of the sea in terms of coefficients calculated from the individual probabilities of specific Fock space giving desired quantum numbers to each baryon. We also used certain modifications, where higher multiplicities are suppressed. The simultaneous effects of these coefficient, on the weak decay matrix elements and semileptonic decays suggest a dominant vector sea, in deciding the properties of these baryons in SU(3)symmetry and its breaking.

Table 3.2 shows the axial-vector coupling constant ratios for $\Delta s = 1, 0$ decays and mention the calculated values in different theoretical models. It can be analyzed from the table 3.2 that the decay ratio for $\Sigma^- \to n$ is showing large error of about 15-16 % from experimental value for both flavor asymmetric valence and sea. This leads to conclusion that this particular type of decay needs higher order corrections in symmetry breaking to be matched with experiments. On the other hand, result for weak decay ratio $\Xi^0 \to \Sigma^+$ agrees well with experimental data with small percentage error of just 1% when both sea and valence is flavor asymmetric. $g_A/g_{V(\Xi^- \to \Sigma^0)}$, $g_A/g_{V(\Xi^- \to \Lambda)}$ and $g_A/g_{V(\Xi^- \to \Xi^0)}$ with breaking in sea, gives data pretty close to theoretical approaches like chiral quark solitan model[27,28], whose results are present in table 3.3. In the operator formalism method, the parameter 'r' becomes more decisive for doubly strange rather than singly strange baryons. Thus it can be stated that statistical model give more accurate predictions for particles having higher mass.

With a flavor asymmetric both valence and sea, the strange quark polarization densities are significantly affected. Spin distribution due to strange quarks in all strange baryon matches with the chiral quark solitan model unlike the light quark spin distribution, either for SU(3) symmetry or its breaking. However, the mismatch between the results does not make any impact on authenticity of our model as large errors can be seen in the results of Ref.[27,28].

A remarkable increase in the value of F/D for broken symmetry in sea is observed. The best fitted values of F and D are 0.48 and 0.71 respectively. The ratio F/D comes out to be 0.676 and found to be deviating about 17% from the experimental value 0.575±0.016 [29]. The extracted values of F and D are further useful to find the total spin distribution of nucleon i.e. $\Gamma_1^p$ and $\Delta\Sigma$ of a nucleonic system. In case of complete SU(3) symmetry, we find $\Delta u = 0.91, \Delta d = -0.24, \Delta s = 0$ respectively. With $m_s$ corrections the spin polarized densities changes to $\Delta u = 0.76, \Delta d = -0.18, \Delta s = -0.019$. This indicates that even in SU(3) symmetry breaking, strange quark contribution to the spin of proton is very small. This small value of strange quark contribution to spin of proton is in favor HAPEX Collaboration [37].

## 1.6 Conclusion:

Our results provide a deeper understanding for baryon structure, with limited quark-gluon Fock states and there by motivating recent experiments to inspect the spin distribution among the quarks and gluons within strange baryons. Present framework suggests a stronger base to choose statistical model with detailed balance principle to verify the experimental and theoretical value available and hence provide a deeper understanding to the strange baryon structure. Symmetry breaking effects lead to reduction in the values of axial matrix elements. Its contribution has a significant role in determining the validity of present approach.

**Acknowledgement**

This work has been partially supported by university grant commission Government of India via SR/ 41-959/2012.